# Deciphering intrinsic inter-subunit couplings that lead to sequential hydrolysis of $F_1$-ATPase ring

**Liqiang Dai[1], Holger Flechsig[2], and Jin Yu[1]***

[1] Beijing Computational Science Research Center, Complex System Research Division, Beijing 100193, China

[2] Department of Mathematical and Life Sciences, Graduate School of Science, Hiroshima University, Hiroshima 739-8526, Japan

*Corresponding author:
Email: jinyu@csrc.ac.cn (JY)

# Abstract

The rotary sequential hydrolysis of metabolic machine $F_1$-ATPase is a prominent feature to reveal high coordination among multiple chemical sites on the stator $F_1$ ring, which also contributes to tight coupling between the chemical reaction and central γ-shaft rotation. High-speed AFM experiments discovered that the sequential hydrolysis was maintained on the $F_1$ ring even in the absence of the γ rotor. To explore how the intrinsic sequential performance arises, we computationally investigated essential inter-subunit couplings on the hexameric ring of mitochondrial and bacterial $F_1$. We first reproduced the sequential hydrolysis schemes as experimentally detected, by simulating tri-site ATP hydrolysis cycles on the $F_1$ ring upon kinetically imposing inter-subunit couplings to substantially promote the hydrolysis products release. We found that it is key for certain ATP binding and hydrolysis events to trigger or accelerate the neighbor-site ADP and Pi release to support the sequential hydrolysis. The kinetically feasible inter-subunit couplings were then scrutinized through atomistic molecular dynamics simulations as well as coarse-grained simulations, in which we enforced targeted conformational changes for the triggering ATP binding or hydrolysis. Notably, we detected the asymmetrical neighbor-site opening that would facilitate the ADP release upon the enforced ATP binding, and computationally captured the complete Pi release through charge hopping upon the enforced neighbor-site ATP hydrolysis. The ATP-hydrolysis triggered Pi release revealed in current TMD simulation confirms a recent prediction made from statistical analyses of single molecule experimental data in regard to the role the ATP hydrolysis plays. Our studies, therefore, elucidate both the concerted chemical kinetics and underlying structural dynamics of the inter-subunit couplings that lead to the rotary sequential hydrolysis of the $F_1$ stator ring.

# Author Summary

Ring-shaped NTPases usually maintain high coordination among multiple subunits on the ring to generate force or torque onto a protein or DNA substrate threading through the center of the ring. Among them, $F_1$-ATPase is an exemplary high efficiency nano-machine that tightly couples rotatory sequential hydrolysis on the ring with the central γ-shaft rotation. Experiments found that the sequential hydrolysis maintains on the $F_1$ stator ring, however, even without the γ rotor in the center. Accordingly, we employed computational methods to investigate how such sequential coordination arises intrinsically, owning to neighbor subunit interactions or couplings on the ring. Kinetically feasible couplings along with structural dynamics basis were elucidated from the combined stochastic modeling, atomistic, and coarse-grained simulations. In particular, we found asymmetrical site openings on the ring to facilitate ADP and Pi releases upon the neighbor-site ATP binding or hydrolysis. The coupling scenarios highlight molecular design principles to synchronize chemical reactions with protein conformational propagations, which may apply to a group of similar types of ring NTPase motors.

# Introduction

The coordination on ring-shaped NTPases is fundamental and crucial for diverse physiological functions of these cellular molecule machines [1-5]. In well-studied systems such as $F_oF_1$-ATP synthase, the binding change mechanism, for example, had been proposed early to illustrate rotary cooperative catalysis around the hexameric $F_1$-ATP ring [6,7]. The mechanism indicated that binding and catalytic activities of ATP alternate sequentially on three chemically active sites, formed at the interfaces of the α and β subunits on the ring [8]. According to the mechanism, the central γ-shaft across the $F_1$ ring dictates the coordinated alternation [9-12]. Nevertheless, experimental studies later on showed that upon significant truncations of the γ-shaft, the $F_1$ motor still rotates directionally [13-16]. Furthermore, recent experiments conducted in the absence of the central γ-shaft demonstrated that the stator $F_1$-ring also maintains the rotary sequential hydrolysis around the three chemical sites [17]. These studies consistently indicated that the sequential coordination arises intrinsically from the $F_1$-ATPase ring, though the stator-rotor interactions would likely enhance the rotary cooperativity.

In this work, we aimed to investigate how the intrinsic sequential coordination around the $F_1$-ATPase ring arises due to inter-subunit or site couplings, in the absence of the central γ-shaft. We assume that the sequential hydrolysis scheme achieved without the γ rotor is similar to that detected originally with the rotor. We focused on two typical reaction schemes, one suggested recently for mitochondrial $F_1$ (as $MF_1$) [18], and the other determined for bacterial $F_1$ (as $BF_1$) [19]. For an individual ATPase site, formed by one α-β hetero-dimer, the uni-site reaction rates for the product ADP and Pi releases are particularly low [20,21]. To allow sufficiently fast reactions as that achieved on the tri-site $F_1$ ring, it is necessary that the inter-subunit couplings significantly accelerate the respective product releases, but how that happen is elusive. Inter-subunit coordination has been widely recognized for ring-like NTPases [1,2], such as viral RNA or DNA packaging motor [22,23], helicases [4,5], and $F_1$-ATPase [2,6,24]. For example, in viral phi29 DNA packaging motor, it has been discovered that multiple subunits coordinate ADP releases with ATP bindings one after another or in an alternating fashion during a dwell phase [1,23,25], while Pi releases happen from one subunit to the next in a burst phase. Nevertheless, it is not clear what specific inter-subunit interactions support the coordination, and it remains a common puzzle for various ring-shaped NTPases. Some suggestions had been made, for example, addressing that ATP binding couples with neighbor-site ADP release [1,23-25]，arginine-finger insertion from one subunit induces ATP hydrolysis in the next site [22,23,25], ATP hydrolysis may facilitate the Pi release in the neighbor site [26], or the Pi release facilitates the hydrolysis. A basic scenario is that one chemical event triggers certain conformational changes that propagate through the subunit interfaces to the neighbor-site(s), accelerate (or de-accelerate) the site conformational transitions on the chemical reaction path, with both kinetic and structural dynamics consequences. The quantitative studies to reveal these couplings and to determine the kinetics or structural dynamics underneath are nevertheless lack of.

The major task of our study is, therefore, to present quantitative description and physical understanding of the inter-subunit couplings on the $F_1$-ATPase ring that essentially support the sequential hydrolysis. Computational modeling and simulations have contributed significantly to the understanding of molecular mechanisms of $F_1$-ATPase [27-36]. In particular, atomistic molecular dynamics (MD) simulations provide detailed structural dynamics to the system, and the simulation time scale moved gradually from several nanoseconds [28] to tens to hundreds of nanoseconds and above [35]. Nevertheless, detecting micro to millisecond 'long-time' conformational changes in such a protein machine is still too hard for conventional all-atom MD. Hence, to probe the inter-subunit conformational couplings, we enforced or accelerated certain conformational transitions to mimic the triggering chemical events in the conventional MD, and combined the atomistic MD with long-time chemical kinetic and coarse-grained structural approaches.

While most of previous computational studies focused on quantifying the rotary motion of $F_1$-ATPase ($\alpha_3\beta_3\gamma$), current work neglected the rotational degree of freedom but studied only a minimal model of the $F_1$ ring for the sequential hydrolysis ($\alpha_3\beta_3$). We started by conducting stochastic kinetics simulations for the three independent ATP hydrolysis cycles, following the uni-site reaction rates obtained for bovine heart mitochondria $F_1$ [21] and *E. coli* $F_1$ [20], respectively. Meanwhile, inter-site couplings that likely happen among the three active sites of the $F_1$ ring were kinetically incorporated and adjusted in the simulation via tuning the involved kinetic rates. If the sequential hydrolysis emerged as experimentally detected [18,19], we then regard those inter-subunit couplings *kinetically feasible*.

For current practice, we adopted a scenario that Pi releases after the ADP release [19,37]. Through the above quantitative analyses, we were able to determine kinetically feasible inter-subunit couplings that support the sequential hydrolysis. Then, we implemented atomistic MD simulations to probe the physical or structural basis of those couplings. The bovine $MF_1$ structure was employed in the MD simulation. To accelerate the triggering chemical events in the MD due to the limited simulation time scale (at sub-microseconds), we implemented the targeted MD (TMD) to one chemical site on the ring, mimicking the conformational changes of the ATP binding or hydrolysis event, and then monitored mechanical responses or conformational relaxations in the two neighboring active sites. We consequently found asymmetrical neighbor site openings that may facilitate the product(s) release or bias the ATP binding for the directional coordination. To validate the TMD results obtained under enforcements and accelerations, we further implemented targeted elastic network (TEN) simulations in a coarse-grained model to see whether the slow relaxation dynamics (overall micro to milliseconds e.g.) well captured in the coarse-grained model appear consistently with our hundreds of nanosecond TMD simulations.

## Results

### 1. Stochastic simulations identified kinetically feasible inter-subunit couplings that support the sequential hydrolysis on the $F_1$ ring

In order to probe what type of inter-subunit or site couplings that are responsible and kinetically feasible for the sequential hydrolysis of the $F_1$-ATPase ring, we first conducted the kinetic Monte Carlo (KMC) simulations for three ATP hydrolysis cycles to mimic the chemical kinetics of the three active sites (see **Methods**). The three chemical cycles proceed independently when there is no inter-subunit coupling connecting the three sites. The coordinated sequential hydrolysis arises kinetically only when certain rates of one hydrolysis cycle are connected with another hydrolysis cycle via some coupling mechanisms or rules.

We started by simulating the three independent hydrolysis cycles using uni-site reaction rates obtained experimentally [20,21] (see **Supplementary Information SI Table S1 and S2**). For each hydrolysis cycle, it goes through ATP binding (from E, the empty, to T, the ATP-bound state), hydrolysis (T or T* to DP, the ADP+Pi bound state), and product release (DP to P, the Pi-bound state, and then to E), under the uni-site rates. Occasionally when two sites among the three are bound with ATP, we refer to T*-site the one binds ATP tighter or earlier, so that it proceeds to an intermediate T* state ready for catalysis [29], i.e., with a higher ATP hydrolysis rate than that of T, an initially ATP bound state.

We then imposed coupling rules into the three ATP hydrolysis cycles in the KMC simulations (see **SI and Fig S1**), e.g., by tuning the product release/unbinding rate of one site upon a certain chemical event in a neighboring site. Only when these rates were tuned sufficiently large at the right moments, can three hydrolysis cycles become sequentially correlated. For that we referred the coupling kinetically feasible. Specifically, the sequential performance was evaluated by a score function (between 0 and 1) based on experimentally detected sequential schemes [18,19]. A higher score shows a better performance (see **SI Eq S1 and Fig S2-4**). Consistently, in the high performance score case, high correlations between the neighboring sites reveal in their hydrolysis cycles, as one site follows the other closely moving along the chemical reaction path (see **SI Eq S2 and Fig S5).**

Since it is not clear whether the product binding affinities are altered, or only the product release/unbinding rates change upon the chemical triggering event under the inter-subunit couplings, we imposed the coupling rules kinetically in two ways (see schematics in **Fig 1A**): for the *implementation I* we altered both the product unbinding and binding rates for the same amount to keep constant binding affinities; for the *implementation II* we tuned only the product unbinding rates while keeping the binding rates constant, assuming the binding being diffusion limited. Regardless of the way of the implementation, we found that the key couplings that supports the sequential performance are (see **Fig 1** and **SI Fig S6**):

*1.1 ATP binding to the E-DP-T ring facilitates the product release from the DP site.*
Presumably, as the ATP binding (E→T) induces an open to closed transition in the current site, the next/downstream site (in the *counterclockwise* direction viewed from the C-terminal side of $F_1$) can be disturbed and the product (ADP and Pi) release rate can be affected [1,23,24]. Accordingly, in the kinetic simulation(s), we allowed the ATP binding to the E-DP-T ring to enhance both the ADP and Pi releases in the next site. In $MF_1$, above $\sim10^5$ fold of the rate enhancement (see **SI Fig S3**) was applied to both the ADP and Pi release in the next site (from E-DP-T to T-E-T*, see **Fig 1B**), so that to achieve a sufficiently high sequential performance. In $BF_1$, (from E-DP-T to T-P-DP, see **Fig 1B**), the ADP release rate has to be enhanced upon the previous site ATP binding (above $\sim10^3$ fold, see **SI Fig S4**) to allow the sequential performance; the Pi release rate can be enhanced either along with the ADP release upon the previous ATP binding (as in $MF_1$), or until after the next site ATP hydrolysis (see **SI Fig S2B** and **C**), as analyzed below.

*1.2 The Pi release can be facilitated right after the ADP release or upon ATP hydrolysis in the T-P-T\* ring*

It has been recently suggested, according to statistical analyses on time series data from high-speed single molecule experiments, that the ATP hydrolysis plays a role to facilitate the neighbor-site Pi release [26]. We tested if this coupling scenario between the ATP hydrolysis and Pi release could support the sequential hydrolysis in the KMC simulation of $BF_1$. We found that by enhancing the Pi release rate $\sim10^3$ fold upon the ATP hydrolysis in the next site (see **SI Fig S4**), the sequential hydrolysis emerged. Alternatively, the Pi release in $BF_1$ could also be enhanced right after the ADP release as that in $MF_1$ to support the sequential performance.

In addition to the most essential couplings identified above, we found it necessary to add a constraint to the $MF_1$ ring (but not to the $BF_1$ ring) to prohibit the three-ATP bound (3-T) configuration, which was not experimentally detected [17]. In the absence of an explicit constraint, the population of the 3-T was already low in the simulation (e.g. 3~6% in $MF_1$, below ~1% in $BF_1$), yet the sequential performance score was still low in $MF_1$ (~ 0.2 to 0.3). Imposing the constraint to $MF_1$ by reducing the third ATP binding rate for 10 ~ 100 fold (see **SI Fig S1** to **S3**) substantially improved the score (to above ~0.6), and lowered the 3-T population further in $MF_1$ (below ~1%). On the other hand, there is no need to impose the constraint to $BF_1$, the 3-T configuration was simply absent with existing tri-site kinetics on the $BF_1$ ring, and the sequential performance does not require the constraint.

As mentioned earlier, in order to ensure that the early ATP bound T*-site hydrolyzes sufficiently fast and faster than the newly bound T-site, we also allowed a rate enhancement (~10 to 100) in T*→DP above that of T→DP (see **SI Fig S1** to **S4)**. To further probe the physical basis of the kinetically feasible couplings identified above, we performed atomistic MD simulations as illustrated below.

2. **Atomistic MD simulations demonstrated asymmetrical openings of the neighbor site upon current site closing/tightening induced by ATP binding/hydrolysis**

We performed four sets of targeted MD (TMD) simulations [38], mimicking the ATP binding or ATP hydrolysis in one chemical site, and monitored the ensued mechanical responses from the two neighboring chemical sites: i) targeting the ATP binding from E-DP-T to T-DP-T* to examine how the downstream DP-site and the upstream T-site would be affected, with a control TMD simulation to include the central γ-shaft; ii) targeting the ATP hydrolysis from T-P-T* to T-P-DP, to examine how the upstream P-site and the downstream T-site would react; iii) targeting the 1st-ATP binding from E-E-E to E-T-E, to examine the responses without the ligand binding in both neighboring sites; iv) targeting the 3rd-ATP binding to E-T*-T , to see if the 3-T ring remains stable.

Note that in each of the TMD simulations, the protein conformational changes corresponding to the ATP binding or hydrolysis triggering event were enforced to happen to the single chemical site to accelerate the originally slow process. The initial and final conformations of the single targeted site (e.g. T-site for ATP binding and DP-site for hydrolysis) were commonly known. The initial structures of the tri-site ring, however, were often made from existing crystal structures, while the final ring structures were usually unknown but were then generated from our TMD simulations.

*i. ATP binding to E-DP-T opens the downstream DP-site to facilitate the ADP release*
We obtained an E-DP-T ring from the P-DP-T structure of the bovine $MF_1$ (PDB: 1E1R) [39], and performed the TMD simulation to enforce the $\alpha_E$-$\beta_E$ structure to approach to the targeted $\alpha_T$-$\beta_T$ conformation, so that the original E-site was turned into a T-site and a T-DP-T* ring was obtained (see **Methods** and **Fig 2**).

To determine if the ATP binding site transited from the open to closed form as being enforced to, we calculated the C-terminal protrusion (i.e., on the DELSEED motif) height, the hinge-bending angle on the β subunit, and the size of the ATP binding pocket (see **SI** and **Fig S7**). Large values of these measures indicate an open site while small values are for a closed one. In **Fig 2A**, we see that the $\beta_E$-protrusion decreased ~ 6 Å during the TMD simulation, indicating clearly a transition from an open site to a closed one (E→T). In response, the average $\beta_{DP}$-protrusion increased slightly (~ 1 Å; also see **Fig 2B**), the $\beta_{DP}$ hinge-bending angle increased (~ 5˚), along with the small pocket-size increase (~ 0.2 Å; see **SI Fig S8**), all indicating a slight opening of the DP-site. In contrast, the average $\beta_T$-protrusion decreased a little (~ 0.2 Å), the pocket size of the T-site shrunk (~ 0.4 Å), though the hinge-bending angle of $\beta_T$ increased slightly (~ 1˚). Movies were made from the simulation for both the top and side views, highlighting how the β-protrusions shifted (**SI Movie S1a** and **S1b**). Since the average magnitudes of the responses were small, we performed an additional TMD simulation for comparison, including the γ-shaft into the center of the E-DP-T ring. The average $\beta_{DP}$-protrusion then increased to ~ 3 Å upon the E-site closing (see **SI Fig S9**), though the changes from other measures remained similarly as in the absence of the γ-shaft. In brief, all three geometric measures consistently suggested that the downstream DP-site opened slightly in response to the ATP binding in the current E-site, while the upstream (clockwise) neighbor T-site did not show such an opening tendency. In particular, one could see that the DELSEED of $\beta_{E \to T}$ moved toward the center of the ring during the enforced ATP binding transition (see **Fig 2C**). The trend was kept in the presence of the γ-shaft, though the γ-shaft occupancy prevents the $\beta_{E \to T}$ protrusion from interacting directly with that of $\beta_{DP}$, and the $\beta_{DP}$ protrusion then increased largely upon the γ-shaft rotation.

Notably, without the γ-shaft, one still sees that the DP-site interface widens between $α_{DP}$ and $β_{DP}$ upon the enforced ATP binding (see **Fig 2D**). The conformational changes get quite significant in the peripheral region of the DP-site (within 10 Å of the bound ADP), with an RMSD ~ 3 Å upon the enforced ATP binding to the E-site (see **SI Fig S10A**). The average number of hydrogen bonds in that region also decreased from ~25 (in the equilibrium control) to ~20 in the TMD simulation (**SI Fig S10B**). In particular, the hydrogen bonds that connect $α_{DP}$ and $β_{DP}$ and ones on the β-strand of $α_{DP}$ frequently broke in the TMD simulation, but not in the equilibrium control.

To further probe how the DP-site region responded the enforced E-site ATP binding to facilitate the ADP release, we additionally conducted the steered molecular dynamics (SMD) simulations to pull ADP out of the DP-site from the initial E-DP-T ring (*before* the TMD) and from the final T-DP-T* structure (*after* the TMD), respectively. Three directions were chosen for the ADP pulling (see **SI Fig S11**): upward from the top of the DP bound pocket, outward and inward toward the center of the ring, respectively. Then we conducted three SMD simulations for each direction, to both the E-DP-T and T-DP-T* rings *before* and *after* the TMD simulations. From these simulations, we concluded that the enforced E-site closing does impact on the DP-site to make the ADP release comparatively easy, which likely happen along the upward and/or outside pathways (see **SI** text).

Finally we probed how the mechanical responses propagate from the ATP binding E-site to the respective DP- and T-site in the stator $F_1$-ring, without the γ-shaft. We calculated individual residue correlations with the E-site ATP binding residues (αI343, αR373, βG161 to V164, βY345, βA421, βF424 and T425) mainly on the $β_E$ subunit (see **Fig 3A**), by evaluating first the pairwise cross-correlation values for all residues (see **SI Fig S12**) and then for each individual residue away from the ATP binding site, calculating the overall correlation strength between this residue and all those residues within the ATP binding site (see **Methods**). Close to the E-site, the correlations were maintained similarly high across $α_{E->T}$ and $β_{E->T}$, respectively. The correlation was still maintained high as being propagated from $α_{E->T}$ to $β_{DP}$, due largely to a loop region (αQ405 to D411; see **Fig 3C** *left*) on top of $α_E$ that interacts closely with the $β_{DP}$ DELSEED region (βD394 to D400) and a helix underneath (βD383 to M390). On the other hand, the propagation from $β_{E->T}$ to $α_T$ and to $β_T$ was through a comparatively long distance, and the correlation decayed significantly reaching to $β_T$. Since the ADP/ATP binding site mainly locates on the β subunit, it is reasonable to see that the upstream T-site was affected less than the downstream DP-site by the E-site ATP binding.

*ii. ATP hydrolysis in T- P-T\* facilitates the Pi release from the upstream P-site*
In order to mimic ATP hydrolysis in the T-P-T* ring (PDB 1W0J) [40], we enforced the T*-site to a targeted DP-site conformation, obtained from the crystal structure (PDB 1E1R) [39]. Interestingly, one notices that in the crystal structure of T-P-T*, the P-site Pi is located around the β P-loop, a marginally stabilized position according to a recent MD study probing free energy of the Pi release [35]. It was indicated that the doubly charged Pi group has a tightly bound state located ~7 Å distance inside, while the P-loop bound state is a less stabilized intermediate state close to the exit of the Pi release channel.

Since the ring structure T-P-T* is obtained from the bovine MF$_1$, the marginally stabilized positioning of Pi in this structure, by itself, suggests a coupling scenario consistent with our proposal for MF$_1$ (see **SI Fig S1c**): the Pi release from the P-site of T-P-T* (to T-E-T*) has been facilitated right along with the ADP release (from T-DP-T* to T-P-T*), upon the ATP binding in the upstream site (E-DP-T to T-DP-T* see **Fig 1B** top).

Moreover, we captured in the TMD simulation of the T-P-T* ring a complete Pi release event from the P-site, upstream of the enforced hydrolysis T*-site. Within 50 ns of the TMD simulation, the protrusion of the T*-site dropped slightly, showing a tightening of the site key to the ATP hydrolysis (T*→DP). Meanwhile, one noticed an up to ~3 Å rise of the β$_P$ protrusion for the P-site (see **Fig 4A** and **SI Fig S13**), indicating an opening response of the site within 50 ns that consequently led to the Pi release. The finding that the Pi release was facilitated upon the enforced hydrolysis nicely confirmed the prediction from the recent experimental data analyses [26], conducted for BF1 (*thermophilic Bacillus* PS3).

Remarkably, our TMD simulation demonstrated a complete charge-hopping pathway of the Pi release through three arginine residues and one lysine residue (see **Fig 4C and SI Movie S2a** and **S2b**), which has not been revealed before. The Pi group was initially positioned through hydrogen bonding with both the P-loop from the β$_P$ subunit and the first arginine finger residue Arg373 from the α$_P$ subunit. In response to the enforced hydrolysis T*→DP transition, the P-loop from β$_P$ retracted slightly from Pi. Meanwhile, Arg143, the second arginine finger from α$_P$ which frequently switched its side chain in between Arg373 and Glu144, captured Pi after ~ 50 ns of the simulation, so that Pi dissociated from Arg373. In addition, three negatively charged residues from the β$_P$ subunit, Asp195, Glu199 and Glu202, helped to repel Pi away from β$_P$. Later on, Pi was further transferred to Lys196 (~ 140 ns) and Arg161 (~ 180 ns) on α$_P$, and then released fully from the F$_1$ ring at the end of the 200 ns TMD simulation. In contrast, in an equilibrium control simulation of the T-P-T* ring (for 200 ns), Pi was kept stable in association with the P-loop and Arg373, while the Arg143 side chain was trapped with Asp314 most of time (see **SI Movie S2c**).

To understand why the conformational transition T*→DP impacts on the P-site that is upstream to the enforced T*-site in this case, we probed again the residue-wise correlations on the T-P-T* ring during the enforced hydrolysis transition (see **Fig 5**). In particular, we see that the correlation decayed quickly right away from the T*-site. Nevertheless, one notices that the local region of the P-site correlates well with the T*-site residues. A close examination shows that the correlation propagates through a restricted region, from the hydrolysis site on β$_{T*}$, along a beta-sheet structure (αArg164 to Arg171, αIle345 to Leu352, and αLeu369 to Arg373, see **Fig 5C** *right*), and then reaches the P-site on α$_P$-β$_P$.

*iii. The first ATP binding opens the next/downstream E-site more than the upstream E-site*
In order to investigate whether asymmetrical responses arise on the early stage of the reaction pathway, we conducted the TMD simulation to a fully empty ring E-E-E (see **SI** and **Fig S14** and **S15**). As we enforced an E→T transition on one pair of $\alpha_E$-$\beta_E$ in the E-E-E ring, we observed similar trends of mechanical responses as that in the E-DP-T configuration: The downstream E-site opened comparing to the equilibrium control simulation, while the upstream E-site did not (see **Fig 6A** and **SI Fig S15**). Further calculations on the cross-correlations also show an asymmetrical pattern (**SI Fig S16**). Hence, without ligand binding in the neighboring actives sites, the asymmetrical responses toward the ATP binding already exist in the ring.

Indeed, the asymmetrical responses triggered by the first ATP binding to the ring can bring a bias in recruiting the second ATP. We found through the KMC simulation that if the second ATP is recruited to the downstream site with a slight bias (e.g. a rate enhancement ~10), then the reaction intermediate quickly converges to E-T*-T, followed by E-DP-T (see **Fig 6B**), rather than to T-T*-E and then T-DP-E. Once the E-DP-T configuration is reached, the chemical cycles repeat as that demonstrated in **Fig 1B**, for both $MF_1$ and $BF_1$.

*iv. The third ATP binding to the ring destabilizes both neighboring ATP-bound sites*
In the KMC simulation, the 3-T configuration in $BF_1$ was absent due to the coordinated tri-site kinetic, i.e., T-T-T was simply not on the dominant reaction pathway. In contrast, the 3-T configuration in $MF_1$ needs to be explicitly prohibited to support the sequential performance by reducing the third ATP binding rate (> ~10) (see **SI Fig S3A**). It indicated that the third ATP binding is unfavorable. In the MD simulation, however, we enforced a third ATP binding to the E-T*-T ring to probe the mechanical (in-)stabilities. We noticed an abnormal increase of the pocket size in the upstream T-site, even though the $\beta_T$ protrusion dropped (see **SI Fig S17** and **S18**). Notably, the bound ATP from the downstream T*-site released in one of the two TMD simulation trials, though the T*-site did not open significantly. Hence, the enforced third ATP binding appears unfavorable as it triggers significant distortions or even destabilizes the bound ATP. Likely, an E-T*-T ring would either resist the third ATP binding, or convert to T-E-T* (rather than T-T*-E) if a third ATP binding succeeds by chance.

**3. Targeted elastic network (TEN) simulation reproduced the asymmetrical opening of the neighbor sites upon ATP binding to validate the TMD results**
Note that the ATP binding and hydrolysis along with induced coupling processes involve substantial conformational transitions with characteristic times of micro- to milliseconds or even longer, which cannot yet be captured by conventional atomistic MD. In order to check whether the accelerated TMD simulations of hundreds of nanoseconds captured essential motions of the original slow conformational transitions, without introducing obvious artifacts, we conducted coarse-grained simulations of the $F_1$-ATPase ring. The single bead per residue elastic network (EN) description [41] was employed, which has been widely used to describe the slow or micro- to milliseconds functional domain motions in protein machines [42-44]. To make an easy comparison, we have implemented targeted dynamic schemes as well within the EN model. Conformational changes inside the ring, which corresponded to relaxation processes of the protein elastic network (see **Methods** and **SI**) [45-47], were followed and the neighbor sites were monitored. In particular, the forced E→T transition starting from the E-DP-T ring, and that starting from the E-E-E ring, was investigated at the coarse grained level. In the E-DP-T to T-DP-T* case, the C-terminal β protrusion of the downstream DP-site increased, indicating an opening of the site, whereas the upstream T-site did not show such an opening (see **SI Fig S19**). In the E-E-E to E-T-E situation, the asymmetric responses found in the TMD simulations were also revealed in the TEN simulations, in which the downstream site showed the opening, whereas the upstream site did not (see **SI Fig S20**). Hence, regarding the conformational responses due to the inter-subunit coupling in the $F_1$ ring, agreement between the coarse-grained model and the atomistic-level accelerated description is obtained.

## Discussion

To reveal quantitative features and structural dynamics mechanisms supporting high coordination of a prototypical ring-shaped NTPase, we computationally investigated the intrinsic inter-subunit couplings that lead to sequential hydrolysis of the $F_1$-ATPase ring, in the absence of the central γ rotor. We first explored which types of inter-subunit couplings are kinetically feasible to allow the sequential hydrolysis of the three active sites on the $F_1$-ring through the stochastic simulations. We imposed the couplings to the three ATP hydrolysis cycles by essentially enhancing the ADP and Pi release rates in one cycle/site upon certain neighbor-site ATP binding or hydrolysis event in another cycle. We found that to support the sequential hydrolysis in $MF_1$, the ADP release needs to be accelerated for above ~$10^5$ fold upon the upstream-site ATP binding in the E-DP-T ring, and similarly for the followed Pi release from the same site; in $BF_1$, on the other hand, the ADP release needs to be accelerated for above ~$10^3$ fold, while the Pi release can be accelerated either along with the ADP release, and/or upon the next-site ATP hydrolysis in the T-P-T* ring. In order to elucidate the physical and structural basis of these kinetically feasible couplings, we next performed atomistic TMD simulations and monitored the ensured mechanical responses on the ring, which followed the enforced conformational transition of the ATP binding or hydrolysis to one chemical site on the ring. We found essentially these features below from the combined stochastic kinetic and structural dynamics simulations.

*i. The primary inter-subunit coupling that supports the sequential hydrolysis with a directional bias is the ATP binding facilitated downstream-site product release, arising from the asymmetrical site opening inherent to the architecture of the trimer-of-dimer ring*

Consistently, we found in the atomistic simulation of the E-DP-T ring that the downstream DP-site opens in response to the enforced E-site closing that mimics the ATP binding. The site opening was partially characterized by slight increasing of the β-protrusion, the hinge bending angle, and the pocket size of the binding. Further examinations show notable conformational changes along with weakening of the hydrogen bond network in the peripheral region of the DP-site. All these allosteric impacts induced by the closing E-site could possibly facilitate the ADP release from the DP-site. When the central γ rotor was included in the TMD simulation, the $β_{DP}$-protrusion increase would be easily enhanced. Close examination on the cross-correlation of the E-DP-T ring, in the absence of the γ rotor, shows that the ATP binding induced conformational changes propagate asymmetrically from the E-site to the DP- and T-site, via the $α_E$-$β_{DP}$ and $β_E$-$α_T$-$β_T$ paths, respectively. The asymmetrical downstream site opening is also preserved upon the induced site closing that mimics the first ATP binding to the empty E-E-E ring. The asymmetrical responses thus appear to be *inherent to the architecture of the trimer-of-heterodimer (α-β) ring, in particular, as α and β subunits differ and alternate around the ring while the chemical active site largely locates on β*. The inclusion of the γ rotor into the center of the ring likely enhance the asymmetrical responses, which either promote the ADP release (and initial Pi release) or bias the next ATP binding to downstream of the current ATP binding site. In particular, the bias for the next or second ATP binding downstream allows the sequential hydrolysis to proceed *directionally* (counterclockwise viewed from the C-terminal side).

*ii. The sequential hydrolysis is also supported by proper timing and acceleration of Pi release, which either quickly follows the ADP release upon the upstream-site ATP binding, or later upon the downstream-site ATP hydrolysis via minor conformational propagation; the Pi release was then computationally predicted to proceed in charge hopping.*

On the other hand, the Pi release can be facilitated via two different mechanisms (see **Fig 1B**), either due to a persistent impact from the upstream-site ATP binding that induced the ADP release ($MF_1$ [18]), or due to the ATP hydrolysis in the downstream T*-site ($BF_1$ [19]). The straightforward support of the ATP binding facilitated Pi release comes from the bovine heart $MF_1$ structure (T-P-T*) *per se* (PDB 1W0J) [40], in which the Pi group in the P-site (after the ADP release) is located nearby the P-loop, a marginally stabilized state, rather than being located in the innermost tightly bound state [35]. The ATP binding to the ring (from E-DP-T to T-DP-T*) not only destabilized ADP from the DP-site, but also Pi from the same site, so that Pi moved from the tightly bound state to the marginally stabilized state not far from exit of the release channel. The release of Pi thereafter then becomes comparatively easy.

Nevertheless, our equilibrium MD simulation of the T-P-T* ring was not yet able to capture the Pi release from the P-site within hundreds of nanoseconds. In the equilibrium simulation, the Pi group was associated with the β P-loop as well as the arginine finger αArg373. The other arginine finger αArg143 was trapped with αAsp314 most of time. Nicely, in contrast, a complete Pi release was predicted computationally in the TMD simulation, as the P-site mechanistically responded to the enforced transition mimicking the ATP hydrolysis in the T*-site. The P-loop retracted and Arg143 was able to switch occasionally toward Arg373, while the three negatively charged residues from $β_P$ also repelled Pi to assist the release. Consequently, the chance that Pi dissociates from the P-loop to be captured by Arg143 increases significantly, which becomes critical for further charge hopping and the final release of Pi. Mutation of any of residues involved in the process would considerably impact on the Pi release.

Comparing with the significant open-to-close conformational changes upon the ATP binding, the ATP hydrolysis induced conformational changes are rather subtle. To see how the subtle conformational changes propagate from the hydrolysis T*-site to the upstream P-site, we found that a restricted beta-sheet region bridged the mechanistic pathway from the T*-site to the P-site. By mutating the key residues for the mechanistic propagation, one would expect the hydrolysis induced Pi release to be hindered. In brief, the ATP hydrolysis does not generate the global responses of the ring as that upon the ATP binding; rather, it triggers restricted local conformational propagation. Besides, in the T-P-T* ring, the P-site upstream to the T*-site is more open and likely much more flexible than the downstream T-site, hence, it easier for the mechanical responses to propagate upstream than downstream.

*iii. The sequential scheme either explicitly requires the third ATP binding inhibited ($MF_1$) or the three-ATP bound configuration was simply not on in the corresponding kinetic pathway in the sequential hydrolysis ($BF_1$)*

Besides, we found additional couplings indispensible for the sequential performances. First, a mechanism aside from the existing tri-site kinetics in $MF_1$ (but not $BF_1$) ring is needed to prevent three active sites from binding ATP simultaneously. Our chemical kinetic simulations showed that the 3-T configuration in both $MF_1$ and $BF_1$ is of low population in the coordinated hydrolysis cycle, even without the mechanical constraint. Nevertheless, the constraint appears to be required further for $MF_1$ to achieve the sequential hydrolysis. Consistently, the TMD simulation enforcing a third ATP binding to the E-T*-T ring demonstrated significant destabilizations to the ring, as the upstream site abnormally enlarged and the downstream site released the bound ATP.

In addition, we adopted a way to enhance the ATP hydrolysis rate in the T*-site, effectively, as a second ATP binds to the ring to form a T-site. Basically, the uni-site hydrolysis rate of $F_1$ is not sufficiently high to allow for the detected tri-site kinetics. The rate enhancement in the hydrolysis-ready intermediate T* ensures that the site binds ATP early almost always hydrolyzes early.

The rotary sequential hydrolysis of $F_1$-ring is constantly supported by chemical free energy through the three ATP cycles, which are coordinated with certain phase differences. The rate enhancements under the inter-subunit couplings correspond to lowering the free energy barriers for the facilitated transitions, which are feasible under two ways (or mixed of both): Either a constant binding affinity or free energy of the ligand (ADP or Pi) to the induced site is maintained, or a constant ligand binding rate is kept. To enhance the ADP release rate ~ $10^5$ fold (in $MF_1$), for example, while keeping a fairly constant ADP rebinding rate, it requires above ~12 $k_BT$ to destabilize the DP-site, with the energy compensation coming from the neighbor-site ATP binding. Since the chemical free energy is mostly dissipated through the rotational degree of freedom in the presence of the γ rotor [48,49], one expects then that the γ rotor plays a significant role to tightly couple the ATP binding to the next-site ADP release, both structurally and energetically. Indeed, our studies show that the $\beta_{DP}$-protrusion responds to the enforced ATP binding much more significantly in the presence than in the absence of the γ rotor. In comparison, the ATP hydrolysis facilitated Pi release requires less energetic compensation or coupling, if there is any, so we could capture the Pi release event notably in the absence of the γ rotor. As such, we infer that the central γ rotor dictates the inter-subunit coupling more significantly in the ATP binding step than in the ATP hydrolysis step, which leads to stronger inter-subunit couplings or higher coordination in the original $F_1$-ATPase than in the stator ring without the rotor. On the other hand, considering the recent MD studies on the γ-shaft rotation and the Pi release [35] together with our simulation results, we suggest that the initial stage of the Pi can also be facilitated by the γ-shaft rotation that accompanies the ATP binding and the facilitated ADP release; further Pi release is then promoted by the neighbor-site ATP hydrolysis in bacterial $F_1$; the complete Pi release couples closely with the further γ-shaft rotation. Based on current work, it would be promising to determine how exactly the central γ rotor contributes to the $F_1$ inter-subunit coordination in the respective steps of the ATP binding and hydrolysis in further experimental and computational studies.

## Conclusion

Combining stochastic simulations on tri-site chemical kinetic of the $F_1$ stator ring with targeted atomistic MD and coarse-grained EN simulations, we inferred the most crucial inter-subunit couplings leading to the sequential hydrolysis on the $F_1$ ring: The ATP binding facilitated downstream ADP release along with the Pi release, as well as the ATP hydrolysis facilitated upstream Pi release. The couplings come from asymmetrical neighbor site opening/loosening responses upon current site closing/tightening during the ATP binding/hydrolysis. The dominant conformational asymmetry, e.g., triggered upon the ATP binding, appears embedded in the trimer-of-heterodimer ring, owning largely to the alternating arrangement of α and β subunits. Alternatively, minor conformational asymmetry reveals upon subtle conformational transitions of the ATP hydrolysis to allow the facilitated Pi release. The mechanistic propagation around the ring, however, is also modulated by the ligand (ATP, ADP or Pi) binding to the neighboring sites. Additionally, preventing a third ATP from binding to the ring, as well as allowing a hydrolysis-ready intermediate to have an elevated hydrolysis rate, also enable the sequential hydrolysis on the $F_1$ ring. Whether comparable inter-subunit coupling scenarios arise for evolutionarily connected protein enzymes with similar molecular architectures would be of high interest to pursue further.

## Methods

### 1. Kinetic Monte Carlo simulation of three chemical sites

For each ATP binding or catalytic site at the α-β interface, there are several chemical states: the empty state without any substrate binding (E); the ATP bound state and the hydrolysis ready state (T/T*); the post-hydrolysis state right after the reaction as ATP turns into ADP and Pi (DP); and the ADP released but Pi bound state (P). Here we follow the convention that the ADP releases prior to the Pi release [19,37]. As such, the uni-site reaction proceeds in cycles as E⇔T/T*⇔DP⇔P⇔E. We differentiate T* from T by enhancing the ATP hydrolysis rate of the T*-state above that of the T-state.

The unite-site reaction rates for forward and backward transitions in the bovine heart mitochondrial and bacterial *E. coli* F1-ATPase were obtained experimentally [20,21] (see **SI Table S1** and **S2**). We assume that the rate of the ADP release from the DP state is the same as that originally listed from the D state, and the rate of the Pi release from the P state is the same as that originally listed from the DP state. We also assume that the bovine and human mitochondrial F1 behave similarly [18], while *E. coli* and *thermophilic* (*Bacillus PS3*) bacterial F1 perform alike as well [50].

We used the kinetic Monte Carlo (KMC) method [51] to simulate the chemical kinetics of $F_1$-ATPase hydrolysis, from the three independent uni-site reactions to the coupled tri-site reactions. In the KMC simulation, one starts with an initial ring configuration (e.g. E-E-E or E-DP-T) of the three chemical sites; each time a transition from one of the three chemical sites was made, either forward or backward along the reaction path, according to rate competitions of all potential transitions in the current ring configuration (i.e., fast transitions get higher chances). When the uni-site reaction rates were adopted for each site, the three sites perform as if they hydrolyze independently. In any coupled tri-site reaction scheme, inter-subunit couplings were imposed at certain ring configurations by tuning certain chemical transition rates upon some neighbor site transitions (e.g. ATP binding E→T). In particular, we used four parameters to define the strength of the couplings, and we tuned these parameters for optimal sequential performances (see **SI**).

2. **Atomistic targeted molecular dynamics simulations**

All the MD simulations were performed using the NAMD 3.0 software [52] with CHARMM22 force field for protein and CHARMM27 force field for lipid [53,54], except for the construction of E-DP-T structure using GROMACS [55]. The $F_1$-ATPase structures were obtained from the existing crystal structures: The E-DP-T ring was made from the crystal structure of P-DP-T (PDB: 1E1R) [39]; the γ-shaft, when included, was rotated for ~40° accordingly. An E-T-T* structure was obtained directly from the crystal structure (PDB: 2JDI) [56]. An E-E-E structure was made from E-T-T, and a T-P-T* structure was obtained from the crystal structure (PDB: 1W0J) [40] (see detail in **SI**).

The PDB structures were then solvated with TIP3P water in a cubic box and the minimum distance from the protein to the wall was 15 Å. We neutralized the systems with $Na^+$ and $Cl^-$ to the concentration of 0.15 M. The temperature was set to 310 K and the pressure was 1 bar. The van der Waals (vdW) and short-range electrostatic interactions used a cutoff of 12 Å. The particle-mesh Ewald method was applied to deal with the long-range electrostatic interactions [57]. The solvated system was minimized with the steepest-descent algorithm, followed by 10 ps MD simulation under the canonical ensemble with time-step 1 fs. Then 5 ns equilibrium simulation was performed under the NVT ensemble with a time step of 1 fs, and position restraints on the heavy atoms of protein were imposed during the simulation. After the constrained simulation, the unconstrained equilibrium simulations or the targeted MD (TMD) simulations were carried out under the NPT ensemble with a time step of 2 fs. During the TMD simulation, the structure was constantly enforced toward a moving target conformation (according to the heavy-atom RMSD value), geometrically designated on the reaction path toward a final target structure [38]. Further implementations are found in **SI**.

3. **Performing the elastic-network relaxation dynamics**

The elastic network of the ring was obtained by first replacing each amino acid residue in the atomic structure by a single bead, which was placed at the position of the alpha-carbon atom of the respective residue. These equilibrium positions were denoted by $\vec{R}_i^{(0)}$ for bead $i$. Then, to determine the pattern of network connections, the distances $d_{ij}^{(0)} = |\vec{R}_i^{(0)} - \vec{R}_j^{(0)}|$ between equilibrium positions of any two beads $i$ and $j$ were compared with a prescribed interaction radius $r_{int} = 10\text{Å}$. If this distance was below $r_{int}$, the two beads were connected by a deformable elastic spring with natural length $d_{ij}^{(0)}$. The network connectivity was stored in matrix $\mathbf{A}$ with entries $A_{ij} = 1$, if beads $i$ and $j$ are connected and $A_{ij} = 0$, else. The total elastic energy of the network is $U = \sum_{i,j;i<j}^{N} \kappa \frac{A_{ij}}{2}\left(d_{ij} - d_{ij}^{(0)}\right)^2$. Here, $N$ is the number of beads in the protein network, $d_{ij} = |\vec{R}_i - \vec{R}_j|$ is the actual length of a spring connecting beads $i$ and $j$ in some deformed network conformation, with $\vec{R}_i$ being the actual position vector of bead $i$, and $d_{ij}^{(0)}$ is the corresponding natural length. The spring stiffness constant $\kappa$ is assumed to be the same for all network springs.

When thermal fluctuations and hydro-dynamical interactions are neglected, the dynamics of the network can be described by a set of Newton's equations in the over-damped limit [45-47]. For bead $i$ the equation of motion is

$$\gamma \frac{d}{dt}\vec{R}_i = -\frac{\partial}{\partial \vec{R}_i}U. \qquad (1)$$

On the left hand side of Eq. (1), $\gamma$ is the friction coefficient assumed to be equal for all network beads. On the right hand side are the elastic forces exerted by network springs of beads which are connected to bead $i$. Explicitly, the equations read

$$\frac{d}{dt}\vec{R}_i = -\sum_j^N A_{ij} \frac{d_{ij} - d_{ij}^{(0)}}{d_{ij}}\left(\vec{R}_i - \vec{R}_j\right) + \sigma \vec{f}_i. \qquad (2)$$

Here, we have removed the dependencies on $\gamma$ and $\kappa$ by an appropriate rescaling of time and, after that, added external forces $\vec{f}_i$ which can act on bead $i$ (if $\sigma = 1$).

In the performed target elastic-network (TEN) simulations external forces were applied only to the beads of one of the catalytic $\alpha\beta$-sites in order to induce the desired nucleotide-induced transition there. Such forces were updated in each integration step and had the form $\vec{f}_{i,n} = \frac{\alpha}{N_s}|RMSD_{n-1} - RMSD^*|\vec{d}_{i,n-1}$, for bead $i$. Here, $\vec{d}_{i,n}$ is the difference vector between positions of bead $i$ in the target and the actual conformation of the forced $\alpha\beta$-subunit, determined after superposition of the subunits at integration step $n$ ($RMSD_n$ is the corresponding root mean square displacement), and $RMSD^* = RMSD_0 - RMSD_0 \frac{n}{S}$, with $S$ being the total number of integration steps. $N_s$ is the number of beads of the forced $\alpha\beta$-subunit and parameters $\alpha = 2,500$ and $S = 400,000$ were chosen in the simulations. The set of equations (2) was numerically integrated to obtain the positions of network beads at all moments in time. In the simulations a first-order scheme with a time-step of 0.1 was used. In particular, we performed TEN to mimic the ATP binding to the E-site of the E-DP-T ring, in comparison with the TMD simulation. Further details are found in **SI**.

4. Individual residue correlations with the enforced ATP binding site
The individual residue correlations with the enforced ATP binding site were calculated to evaluate the overall correlation strength between each residue and the binding pocket.

Firstly, we calculated the pairwise correlations for all the residues. The correlation between each pair of residues is given by $C_{ij} = \frac{<(R_i-<R_i>)\cdot(R_j-<R_j>)>}{\sqrt{<(R_i-<R_i>)^2>\cdot<(R_j-<R_j>)^2>}}$, where $R_i$ and $R_j$ are the position vectors of Cα atom of residue i and j, respectively, and the <..> represents the average over the simulation trajectory. Then we obtained a pairwise NxN correlation matrix, where N is the total number of residues.

Finally, we calculated the correlation between residue i and binding pocket by $C_{ip} = \sum_{j \in pocket} C_{ij}^2$, where the pocket residues include αILE343, αARG373, βGLY161, βLYS162, βTHR163, βVAL164, βTYR345, βALA421, βPHE424 and βTHR425.

**Acknowledgements**
We acknowledge the computational support from the Beijing Computational Science Research Center (CSRC) and Special Program for Applied Research on Super Computation of the NSFC-Guangdong Joint Fund (the second phase).

# References

1. Liu S, Chistol G, Bustamante C (2014) Mechanical Operation and Intersubunit Coordination of Ring-Shaped Molecular Motors: Insights from Single-Molecule Studies. Biophysical Journal 106: 1844-1858.
2. Lino R, Noji H (2013) Intersubunit coordination and cooperativity in ring-shaped NTPases. Curr Opin Struct Biol 23: 229-234.
3. Lyubimov AY, Strycharska M, Berger JM (2011) The nuts and bolts of ring-translocase structure and mechanism. Curr Opin Struct Biol 21: 240-248.
4. Enemark EJ, Joshua-Tor L (2008) On helicases and other motor proteins. Curr Opin Struct Biol 18: 243-257.
5. Singleton MR, Dillingham MS, Wigley DB (2007) Structure and mechanism of helicases and nucleic acid translocases. Annu Rev Biochem 76: 23-50.
6. Boyer PD (1989) A perspective of the binding change mechanism for ATP synthesis. FASEB J 3: 2164-2178.
7. Hutton RL, Boyer PD (1979) Subunit Interaction during Catalysis. Alternating Site Cooperativity of Mitochondrial Adenosine Triphosphatase. J Biol Chem 254: 9990-9993.
8. Lesile AG, Walker JE (2000) Structural model of F1-ATPase and the implications for rotary catalysis. Philos Trans R Soc Lond B Biol Sci 355: 465-471.
9. Boyer PD (1993) The binding change mechanism for ATP synthase - Some probabilities and possibilities. Biochimica et Biophysica Arta 1140: 215-250.
10. Abrahams JP, Leslie AGW, Lutter R, Walker JE (1994) Structure at 2.8 Â resolution of F1-ATPase from bovine heart mitochondria. Nature 370: 621-628.
11. Noji H, Yasuda R, Yoshida M, Kinosita Jr K (1997) Direct observation of the rotation of F1-ATPase. Nature 386: 299-302.
12. Adachi K, Oiwa K, Nishizaka T, Furuike S, Noji H, et al. (2007) Coupling of rotation and catalysis in F(1)-ATPase revealed by single-molecule imaging and manipulation. Cell 130: 309-321.
13. Furuike S, Hossain MD, Maki Y, Adachi K, Suzuki T, et al. (2008) Axle-Less F1-ATPase Rotates in the Correct Direction. Science 319: 955-958.
14. Hossain MD, Furuike S, Maki Y, Adachi K, Suzuki T, et al. (2008) Neither Helix in the Coiled Coil Region of the Axle of F(1)-ATPase Plays a Significant Role in Torque Production. Biophysical Journal 95: 4837-4844.
15. Kohori A, Chiwata R, Hossain Mohammad D, Furuike S, Shiroguchi K, et al. (2011) Torque Generation in F(1)-ATPase Devoid of the Entire Amino-Terminal Helix of the Rotor That Fills Half of the Stator Orifice. Biophysical Journal 101: 188-195.
16. Chiwata R, Kohori A, Kawakami T, Shiroguchi K, Furuike S, et al. (2014) None of the rotor residues of F1-ATPase are essential for torque generation. Biophys J 106: 2166-2174.
17. Uchihashi T, Iino R, Ando T, Noji H (2011) High-speed atomic force microscopy reveals rotary catalysis of rotorless F1-ATPase. Science 333: 755-758.

18. Suzuki T, Tanaka K, Wakabayashi C, Saita E-i, Yoshida M (2014) Chemomechanical coupling of human mitochondrial F1-ATPase motor. Nature chemical biology 10: 930-936.
19. Watanabe R, Noji H (2014) Timing of inorganic phosphate release modulates the catalytic activity of ATP-driven rotary motor protein. Nat Commun 5: 3486.
20. Senior AE (1992) Catalytic sites ofEscherichia coli F1-ATPase. Journal of bioenergetics and biomembranes 24: 479-484.
21. Grubmeyer C, Cross R, Penefsky H (1982) Mechanism of ATP hydrolysis by beef heart mitochondrial ATPase. Rate constants for elementary steps in catalysis at a single site. Journal of Biological Chemistry 257: 12092-12100.
22. Lisal J, TUma R (2005) Cooperative mechanismof RNA packaging motor. The Journal of Biological Chemistry 280: 23157-23164.
23. Yu J, Moffit J, Hetherington CL, Bustamante C, Oster G (2010) Mechanochemistry of a Viral DNA Packaging Motor. Journal of Molecular Biology 400: 186-203.
24. Adolfsen R, Moudrianakis EN (1976) Binding of adenine nucleotides to the purified 13S coupling factor of bacterial oxidative phsphorylation. Arch Biochem Biophys 172: 425-433.
25. Yu J (2014) Coordination and Control Inside Simple Biomolecular Machines. In: Han K, editor. Protein Conformational Dynamics, Advances in Experimental Medicine and Biology: Springer International Publishing Switzerland. pp. 353-383.
26. Li C-B, Ueno H, Watanabe R, Noji H, Komatsuzaki T (2015) ATP hydrolysis assists phosphate release and promotes reaction ordering in F1-ATPase. Nature Communication 6: 10223.
27. Wang H, Oster G (1998) Energy transduction in the F1 motor ofATP synthase. Nature 396: 279-282.
28. Bockmann RA, Grubmuller H (2002) Nanoseconds molecular dynamics simulation of primary mechanical energy transfer steps in F1-ATP synthase. Nature Structural Biology 9: 198-202.
29. Dittrich M, Hayashi S, Schulten K (2003) On the mechanism of ATP hydrolysis in F1-ATPase. Biophysical Journal 85: 2253-2266.
30. Gao YQ, Yang W, Karplus M (2005) The binding change mechanism of F1-ATPase revisited. Cell 123: 195-205.
31. Koga N, Takada S (2006) Folding-based molecular simulations reveal mechanisms of the rotary motor F1−ATPase. PNAS 103: 5367-5372.
32. Gaspard P, Gerritsma E (2007) The stochastic chemomechanics of the F1-ATPase molecular motor. Journal of Theoretical Biology 247: 672-686.
33. Mukherjee S, Warshel A (2012) Electrostatic origin of the mechanochemical rotary mechanism and the catalytic dwell of F1-ATPase. PNAS 108: 20550-20555.
34. Ito Y, Yoshidome T, Matubayasi N, Kinoshita M, Ikeguchi M (2013) Molecular Dynamics Simulations of Yeast F1-ATPase before and after 16° Rotation of the γ Subunit. The Journal of Physical Chemistry B 117: 3298-3307.
35. Okazaki K-i, Hummer G (2013) Phosphate release coupled to rotary motion of F1-ATPase. Proceedings of the National Academy of Sciences 110: 16468-16473.
36. Okazaki K, Hummer G (2015) Elasticity, friction, and pathway of γ-subunit rotation in

FoF1-ATP synthase. PNAS 112: 10720-10725.
37. Watanabe R, Iino R, Noji H (2010) Phosphate release in F1-ATPase catalytic cycle follows ADP release. Nature chemical biology 6: 814-820.
38. Schlitter J, Engels M, Krüger P (1994) Targeted molecular dynamics: a new approach for searching pathways of conformational transitions. Journal of molecular graphics 12: 84-89.
39. Braig K, Menz RI, Montgomery MG, Leslie AGW, Walker JE (2000) Structure of bovine mitochondrial F1-ATPase inhibited by Mg2+ADP and aluminium fluoride. Structure 8: 567-573.
40. Kagawa R, Montgomery MG, Braig K, Leslie AG, Walker JE (2004) The structure of bovine F1−ATPase inhibited by ADP and beryllium fluoride. The EMBO Journal 23: 2734-2744.
41. Bahar I, Atilgan AR, Erman B (1997) Direct evaluation of thermal fluctuations in proteins using a single-parameter harmonic potential Folding and Design 2: 173-181.
42. Bahar I, Lezon TR, Yang L-W, Eyal E (2010) Global Dynamics of Proteins: Bridging Between Structure and Function. Annual review of biophysics 39: 23-42.
43. Cui Q, Bahar I, editors (2005) Normal Mode Analysis: Theory and Applications to Biological and Chemical Systems. London/Boca Raton, FL: Chapman and Hall/CRC
44. Zheng W, Doniach S (2003) A comparative study of motor-protein motions by using a simple elastic-network model. PNAS 100: 13253-13258.
45. Flechsig H, Mikhailov AS (2010) Tracing entire operation cycles of molecular motor hepatitis C virus helicase in structurally resolved dynamical simulations. Proc Natl Acad Sci USA 107: 20875-20880.
46. Togashi Y, Mikhailov AS (2007) Nonlinear relaxation dynamics in elastic networks and design principles of molecular machines. Proc Natl Acad Sci USA 104: 8697-8702.
47. Togashi Y, Yanagida T, Mikhailov AS (2010) Nonlinearity of mechanochemical motions in motor proteins. PLos Computational Biology 6: e1000814.
48. Toyabe S, Okamoto T, Watanabe-Nakayama T, Taketani H, Kudo S, et al. (2010) Nonequilibrium energetics of a single F1-ATPase molecule. Phys Rev Lett 104: 198103.
49. Kawaguchi K, Sasa S, Sagawa T (2014) Nonequilibrium Dissipation-free Transport in F1-ATPase and the Thermodynamic Role of Asymmetric Allosterism. Biophysical Journal 106: 2450-2457.
50. Bilyard T, Nakanishi-Matsui M, Steel BC, Pilizota T, Nord AL, et al. (2013) High-resolution single-molecule characterization of the enzymatic states in Escherichia coli F1-ATPase. Philos Trans R Soc Lond B Biol Sci 368: 20120023.
51. Chatterjee A, Vlachos DG (2007) An overview of spatial microscopic and accelerated kinetic Monte Carlo methods. Journal of Computer-Aided Materials Design 14: 253-308.
52. Nelson MT, Humphrey W, Gursoy A, Dalke A, Kalé LV, et al. (1996) NAMD: a parallel, object-oriented molecular dynamics program. International Journal of High

Performance Computing Applications 10: 251-268.
53. MacKerell AD, Bashford D, Bellott M, Dunbrack R, Evanseck J, et al. (1998) All-atom empirical potential for molecular modeling and dynamics studies of proteins. The journal of physical chemistry B 102: 3586-3616.
54. Foloppe N, MacKerell Jr AD (2000) All‐atom empirical force field for nucleic acids: I. Parameter optimization based on small molecule and condensed phase macromolecular target data. Journal of computational chemistry 21: 86-104.
55. Berendsen HJ, van der Spoel D, van Drunen R (1995) GROMACS: A message-passing parallel molecular dynamics implementation. Computer Physics Communications 91: 43-56.
56. Bowler MW, Montgomery MG, Leslie AGW, Walker JE (2007) Ground State Structure of F1-ATPase from Bovine Heart Mitochondria at 1.9 Å Resolution. Journal of Biological Chemistry 282: 14238-14242.
57. Darden T, York D, Pedersen L (1993) Particle mesh Ewald: An N · log (N) method for Ewald sums in large systems. The Journal of chemical physics 98: 10089-10092.

# Figure Legends

**Fig 1**. Stochastic simulations of tri-site hydrolysis reactions on the $F_1$ ring. (A) The trajectories show the number of ATP hydrolysis cycles for the three chemical sites in bovine MF1, from the independent (yellow traces, with three different shades shown for the three active sites) to the partially coordinated (blue traces with three shades) and to the highly sequential case (red traces with three shades as well). In the sequential case, each site follows its upstream/counterclockwise site closely along the ATP hydrolysis reaction path as being demonstrated experimentally. The sequential score, which quantify the sequential performance of the three active sites (see **SI Eq S1**), increased from 0 (independent) to 0.4 (partially coordinated) and to 0.6 (highly sequential), correspondingly. In the independent case, the uni-site reaction rates from the bovine heart $MF_1$ were used [21]. The ring-shaped structure of a key configuration E-DP-T is shown (E for Empty, T for ATP bound state, DP for ADP*Pi or the hydrolysis state, and P for the Pi bound state), with the most essential inter-subunit coupling illustrated: ATP binding to the E-site facilitates the ADP release from the next DP-site. The coupling is imposed in two ways: *implementation I* and *II* (see text), with the *implementation I* for $MF_1$ trajectories shown here. The trajectories for $MF_1$ and $BF_1$ with both implementations are shown in **SI Fig S6**. (B) The schemes for the essential inter-subunit couplings to achieve the sequential hydrolysis in $MF_1$ (*top*) and $BF_1$ (*bottom*): The ATP binding (E→T) to E-DP-T enhances the ADP release (DP→P), for both $MF_1$ and $BF_1$; in $MF_1$, the Pi release (P→E) is enhanced right after the ADP release upon the ATP binding; in $BF_1$, the Pi release is enhanced either similarly as that in $MF_1$ or later upon the ATP hydrolysis (T*→DP).

**Fig 2**. The TMD simulation enforcing an ATP binding to $\alpha_E$-$\beta_E$ in the E-DP-T ring. (A) In the first 150 ns TMD simulation, the E→T transition was gradually enforced; additional simulations were conducted for relaxation as holding the structure toward the final targeted form. The conformational changes of the chemical sites were monitored by the C-terminal protrusions of the β subunits (green for the E-site, blue for DP, and yellow for T). (Inset) The RMSD of $\alpha_E$-$\beta_E$ with respect to the targeted $\alpha_T$-$\beta_T$. (B) The histograms of the protrusions from $\beta_{DP}$ (blue alike colors) and $\beta_T$ (yellow alike), measured from the first 150 ns (light color) TMD, the second and the third 150 ns simulation (darker blue or yellow colors), respectively (with the average changes indicated by arrows). The side view of the conformational changes on the $\beta_{E\text{->}T}$ and $\beta_{DP}$ C-terminal protrusions were shown, with spheres the DELSEED motif. (C) The top view from the C-terminal side of the $F_1$ ring, for the initial (E-DP-T, *left*) and the enforced configuration (T-DP-T*, *right*). The DELSEED motifs are highlighted in magenta spheres. (D) The widening of the DP-site interface. *Left*: The side view of the E-DP-T structure. *Right*: The interface between $\alpha_{DP}$ and $\beta_{DP}$ around the DP-site before and after the TMD simulation. The *initial* structure is shown in transparent, with $\beta_{DP}$ colored pink, $\alpha_{DP}$ colored yellow, and ADP/Pi colored by atoms. The *final* structure after TMD is colored in blue ($\beta_{DP}$) and light blue ($\alpha_{DP}$).

**Fig 3**. The correlation with the ATP binding pocket residues in the TMD simulation of the E-DP-T ring. (A) The correlation map of the full ring during the enforced E→T transition starting from E-DP-T, colored according to the correlation values (blue/white/red: high/medium/low) between individual residues and the E-site ATP binding residues (in green spheres). (B) The residue-wise correlation from $β_{DP}$ and $β_T$ (blue and gray). (C) An inside view showing the correlation propagation from $α_{E->T}$ to $β_{DP}$ and to $α_{DP}$ (*left*), and from $β_{E->T}$ to $α_T$ and to $β_T$ (*right*). The bound ADP+Pi and ATP at the DP-site and T-site are circled, respectively. The interface region of $α_{E->T}$ to $β_{DP}$ was shown with zoom-in views to the left.

**Fig 4**. The TMD simulation enforcing an ATP hydrolysis transition to the T*-site in the T-P-T* ring. (A) The top views of the ring at an early and a late stage of the simulation (T-site green, P-site blue, and T*-site yellow). (B) The distance between Pi and an inside reference (Cα of αR373) in the TMD and an equilibrium (eq) control simulation of the T-P-T* ring. (C) The full pathway of the Pi release captured in the TMD simulation. The Pi group (spheres) hopped from the arginine finger αR373 to αR143 (~ 80 ns), and then to αK196 (~ 140 ns) and αR161 (~ 180 ns) before the final release. Three negatively charged residues from $β_P$ (D195, E199 and E202) also help the Pi release through repulsion. $β_P$ is colored in ice blue and its P-loop in magenta, $α_P$ in cyan. The key charged residues are colored in blue (positive) and red (negative).

**Fig 5**. The correlation with the ATP hydrolysis pocket in the TMD simulation of the T-P-T* ring. (A) The correlation map of the full ring during the enforced T*→DP transition starting from T-P-T*, colored according to the correlation values (blue/white/red: high/medium/low) between individual residues and the T*-site ATP binding/hydrolysis residues (in green spheres). (B) The residue-wise correlation from $β_P$ and $β_T$ (blue and gray). (C) An inside side view showing the correlation propagation from $α_{T*→DP}$ to $β_T$ and to $α_T$ (*left*), and from $β_{T*→DP}$ to $α_P$ and to $β_P$ (right). To understand why the conformational transition T*→DP impacts on the P-site that is upstream to the enforced T*-site in this case, we calculated again the residue-wise correlations on the T-P-T* ring during the enforced hydrolysis transition (see **Fig 5**). In particular, we see that the correlation decayed quickly right away from the T*-site. Nevertheless, one notices that the local region of the P-site correlates well with the T*-site residues. A close examination shows that the correlation propagates through a restricted region, from the hydrolysis site on $β_{T*}$, along a beta-sheet structure (αArg164 to Arg171, αIle345 to Leu352, and αLeu369 to Arg373, see **Fig 5C** *right*), and then reaches the P-site on $α_P$-$β_P$.

**Fig 6**. The TMD simulation enforcing an E→T transition to one pair of $α_E$-$β_E$ in an empty E-E-E ring. (A) The E-site being forced into a T-site is colored blue, the next/downstream E-site colored yellow, and the previous/upstream E-site green (*right*). The β-protrusions from both neighboring E-sites, sampled from the TMD simulation (dark) and a control equilibrium simulation (light), was shown, with arrows indicating the opening or closing

trend (*left*). (B) An early reaction pathway starting from an empty $F_1$-ring (E-E-E) shows a bias after the first ATP binding. The bias likely facilitates the second ATP binding to the downstream site, and further leads to the E-DP-T configuration (instead of T-DP-E).

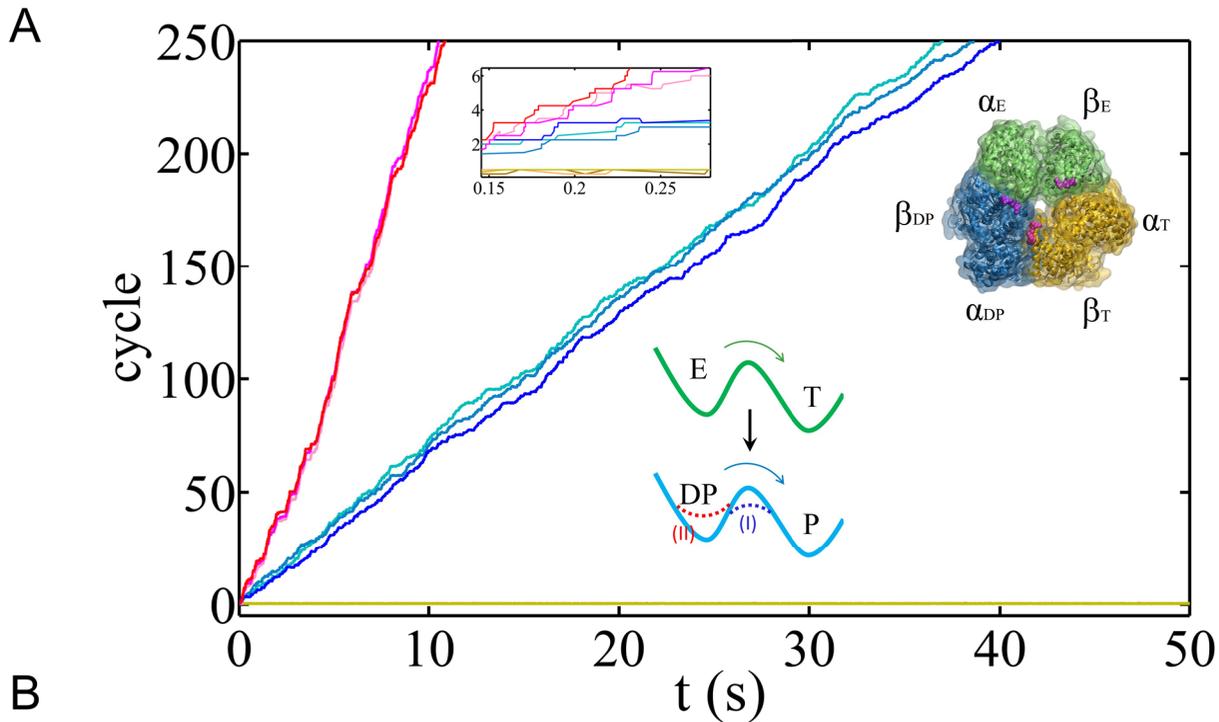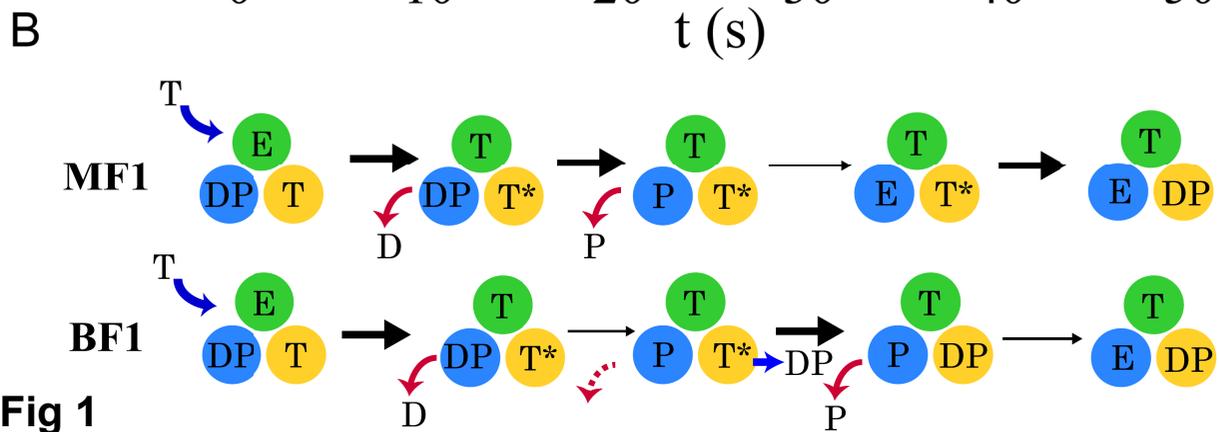

Fig 1

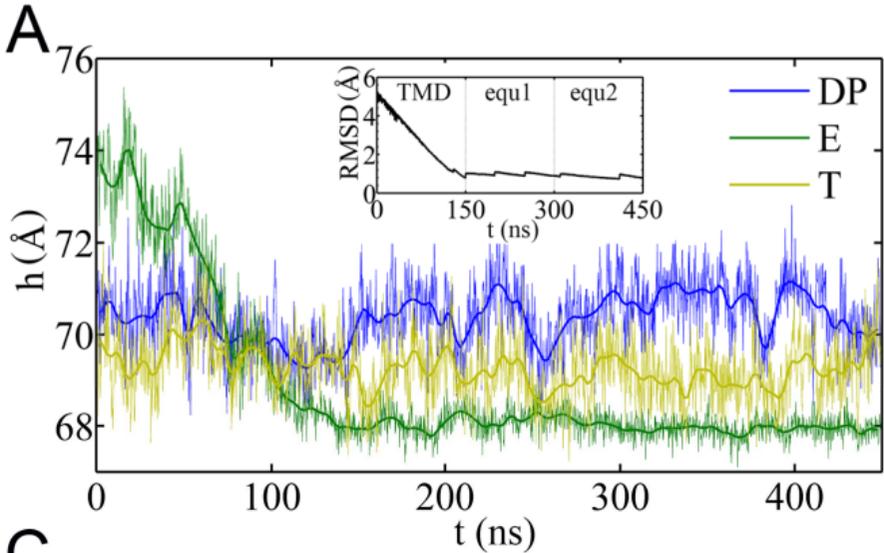
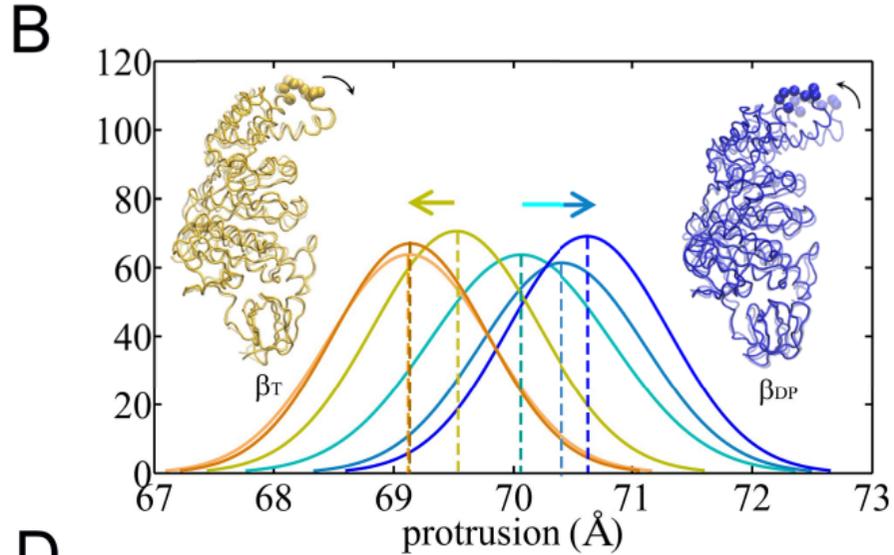
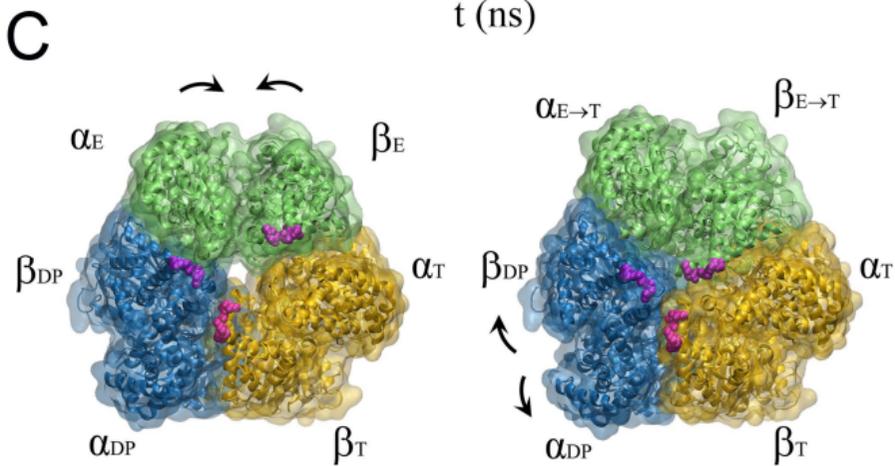
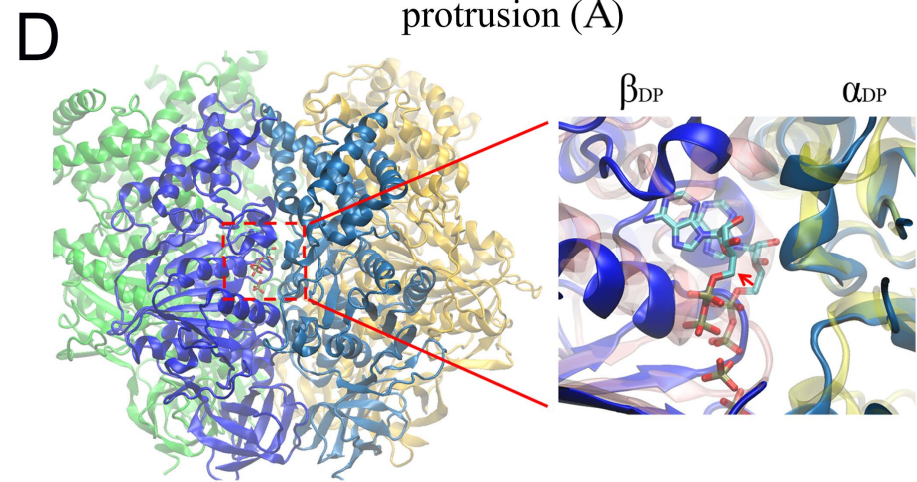

**Fig 2**

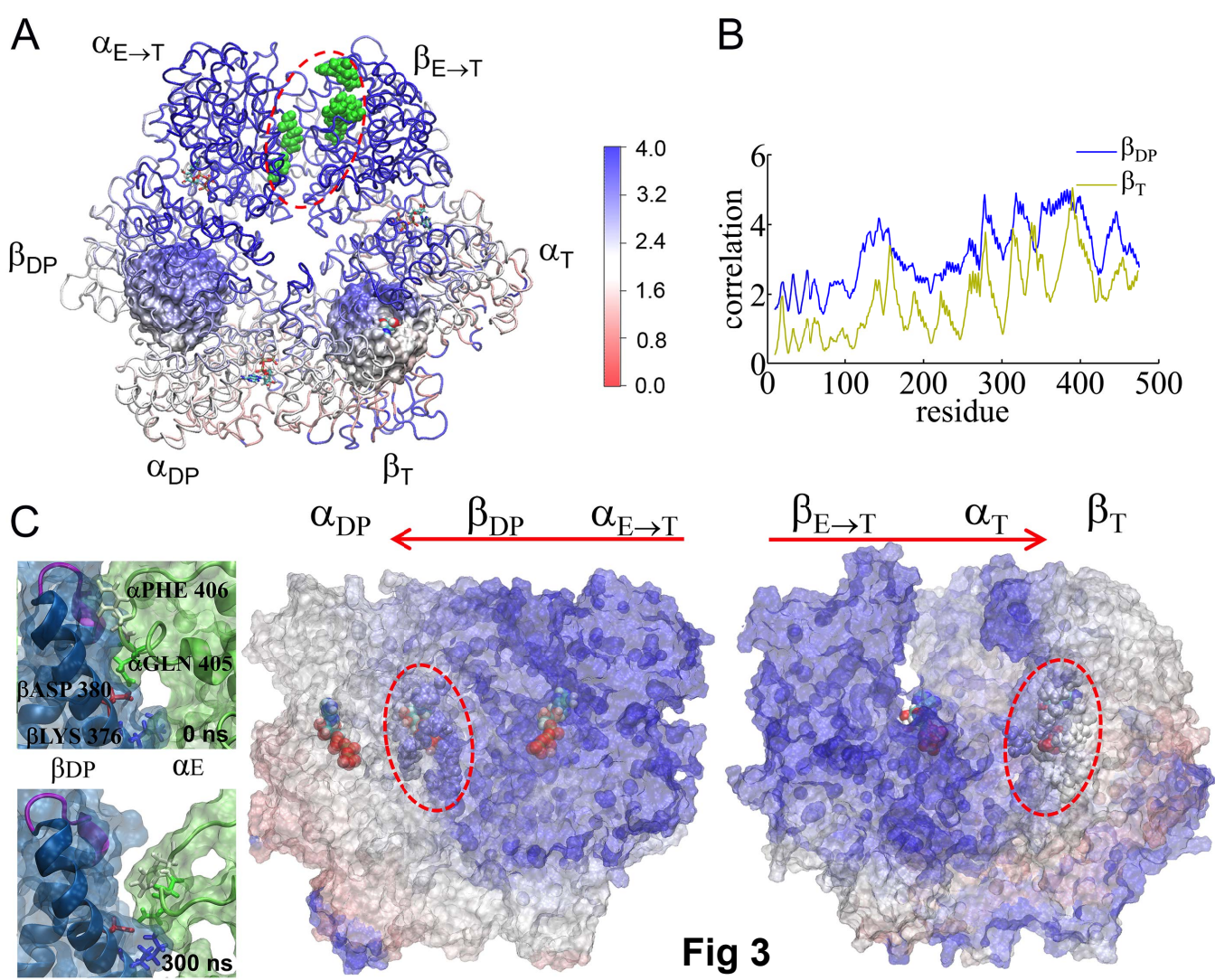

Fig 3

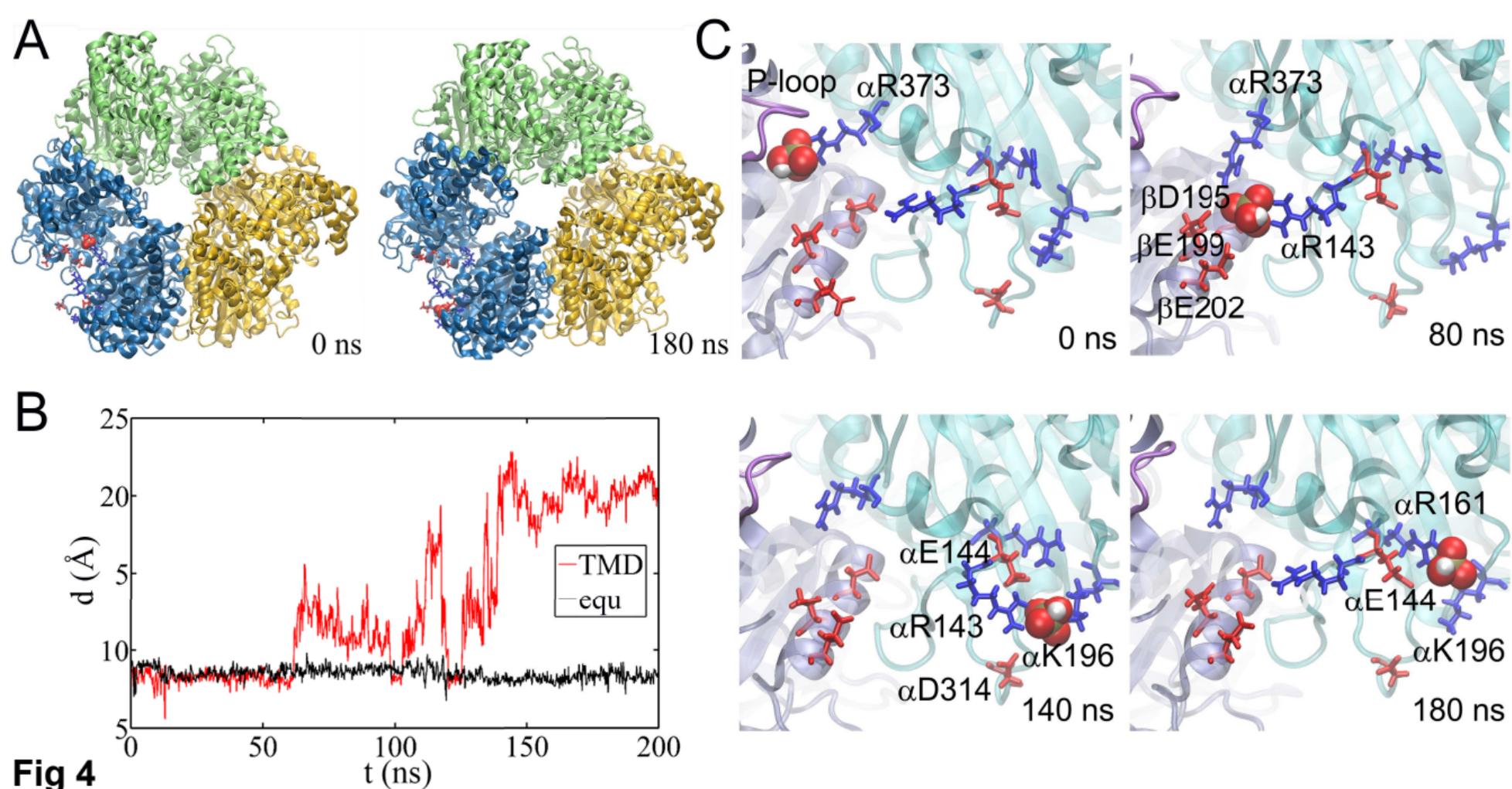

**Fig 4**

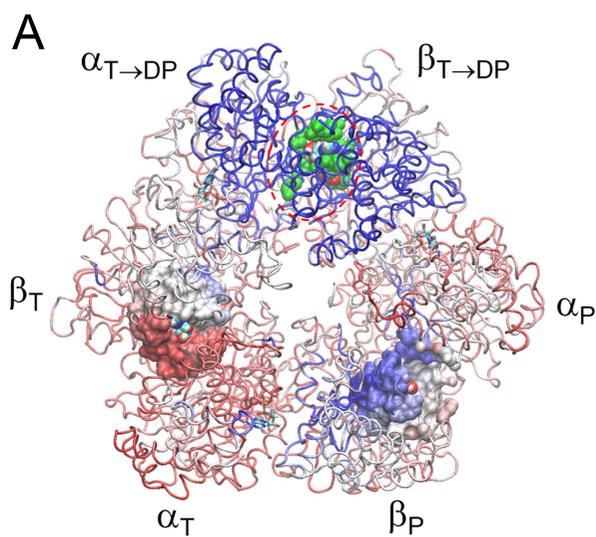
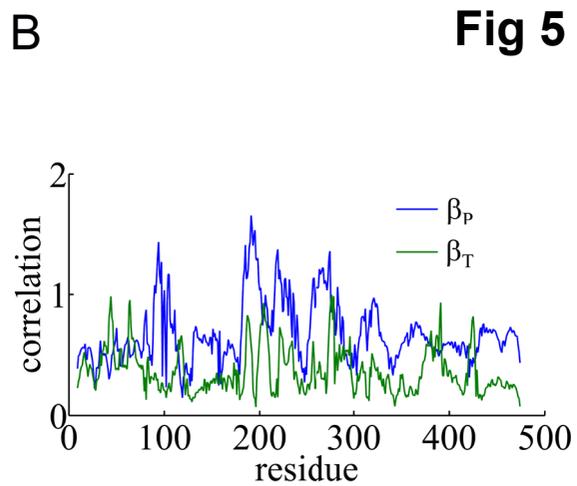
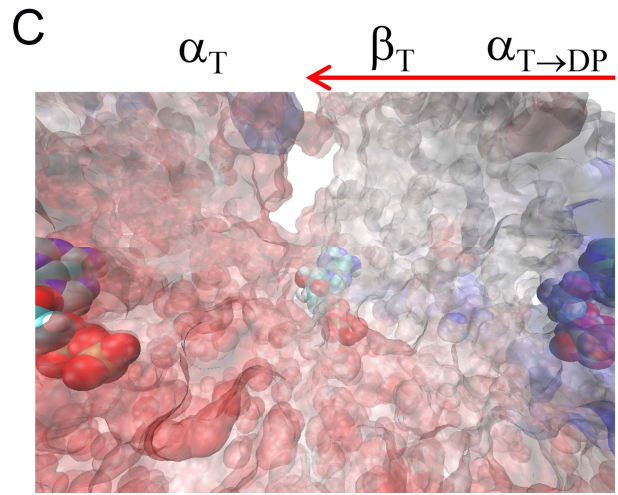
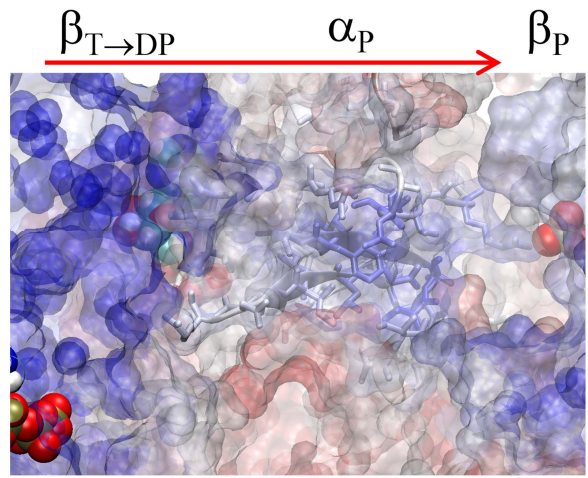

Fig 5

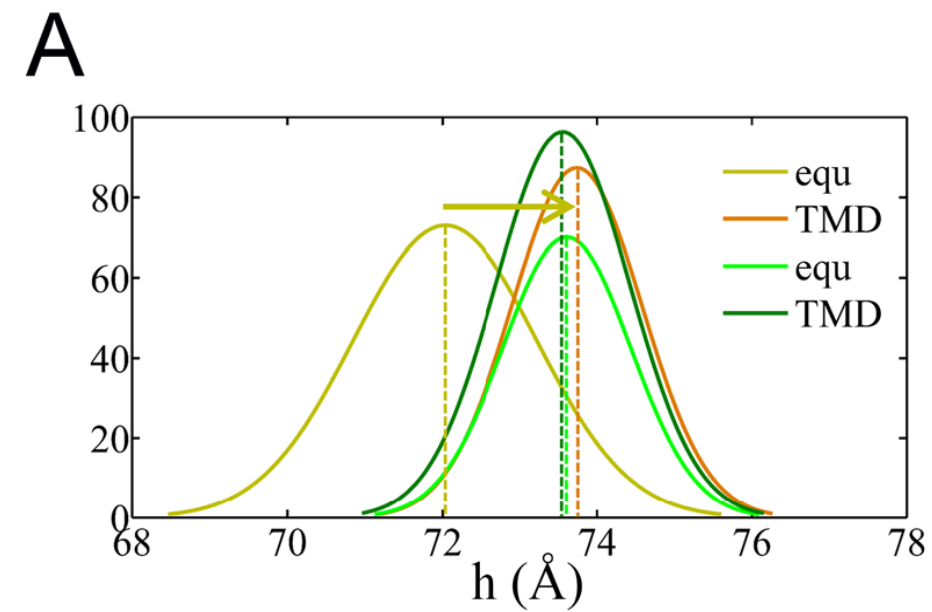
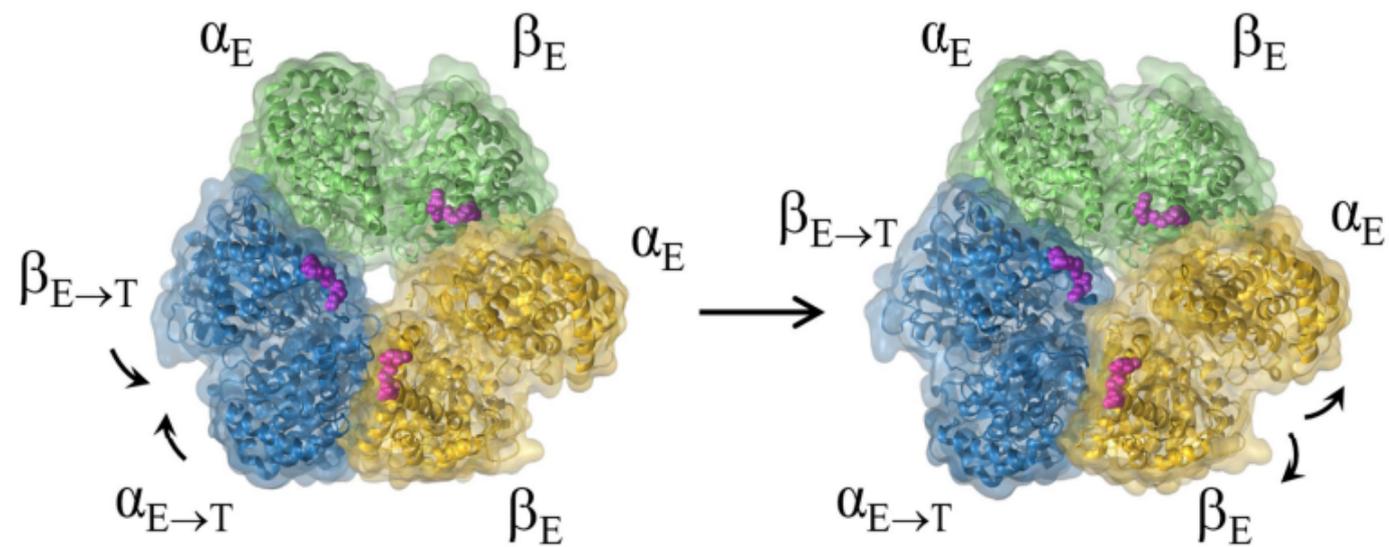
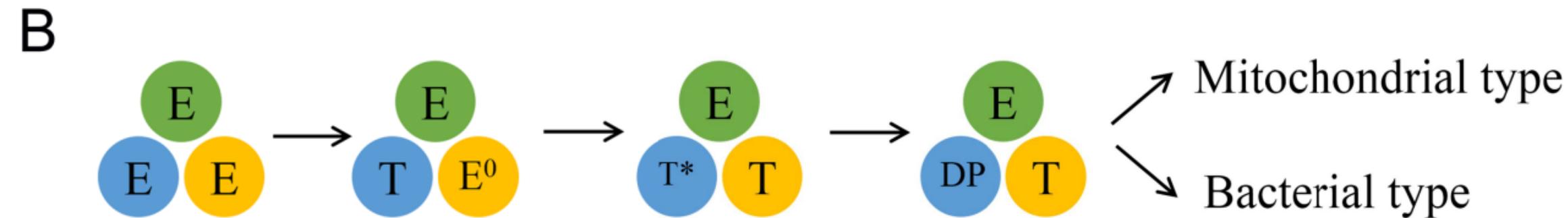

**Fig 6**